\documentclass[oneside, a4paper,12pt,psamsfonts]{amsart}
\usepackage{array,latexsym}
\usepackage[pdftex]{graphicx}
\pdfoutput=1
\usepackage{amsfonts}
\usepackage{amsmath}
\usepackage{amsthm}
\usepackage{a4wide}
\newtheorem{theorem}{Theorem}[section]
\newtheorem{Q}[theorem]{Question}
\theoremstyle{remark}

\begin{document}

\title{Estimation of Bond Percolation Thresholds on the Archimedean Lattices} 
\author{Robert Parviainen}
\address{ARC Centre of Excellence for Mathematics 
   	and Statistics of Complex Systems\\
  	139 Barry Street, The University of Melbourne, Victoria, 3010}
\email{robertp@ms.unimelb.edu.au}
\date{\today}

\begin{abstract}
  We give accurate estimates for the bond percolation critical
  probabilities on seven Archimedean lattices, for which the critical
  probabilities are unknown, using an algorithm of Newman and Ziff. 
\end{abstract}

\maketitle

 \section{Introduction}
Since the introduction of percolation theory, it has been an interesting and
challenging problem to determine the percolation thresholds.  Only
a few non-trivial graphs are exactly solved, such as the bond model on
the square, triangular and hexagonal lattices, and the site model on
the triangular and Kagom\'e lattices. Recently Scullard and Ziff have found many new thresholds for various classes of lattices \cite{S06, Z06}. They also conjecture, \cite{ZS06}, the value of the threshold for one lattice considered here: the $(3,12^{2})$ lattice, and also for the Kagom\'e lattice.
Precise estimates have been calculated for example for the site model on
the Archimedean lattices, \cite{SZ99}, and for the bond model on the Kagom\'e
lattice, \cite{ZS97}. 

Bounds, more or less tight, have been found by various
authors for some lattices. For the bond model on the Archimedean
lattices, see \cite{WP02} for a review of rigorous bounds on
critical probabilities. Recently, Riordan and Walters, \cite{RW07}, have given tight rigorous confidence intervals for both site and bond percolation on all Archimedean lattices.

In this paper, we provide precise estimates for the bond percolation
thresholds for the unsolved Archimedean lattices. 

\subsection{Archimedean lattices}
The Archimedean lattices are the vertex transitive graphs that can be
embedded in the plane such that every face is a regular polygon. A
polygon is regular if all edges have the same length, and all interior
angles are the same. Kepler, \cite{Kepler}, showed that there exists exactly
11 such graphs.

The lattices are given names according to the sizes (number of sides of the polygon) of faces incident
to a given vertex. The face sizes are listed in order, starting with
a face such that the list is the smallest possible in lexicographical
order. The square lattice thus gets the name $(4,4,4,4)$, abbreviated
to $(4^4)$, and the Kagom\'e lattice the name $(3,6,3,6)$.

\begin{figure}[htbp]
  \begin{center}
    \includegraphics[scale=0.65]{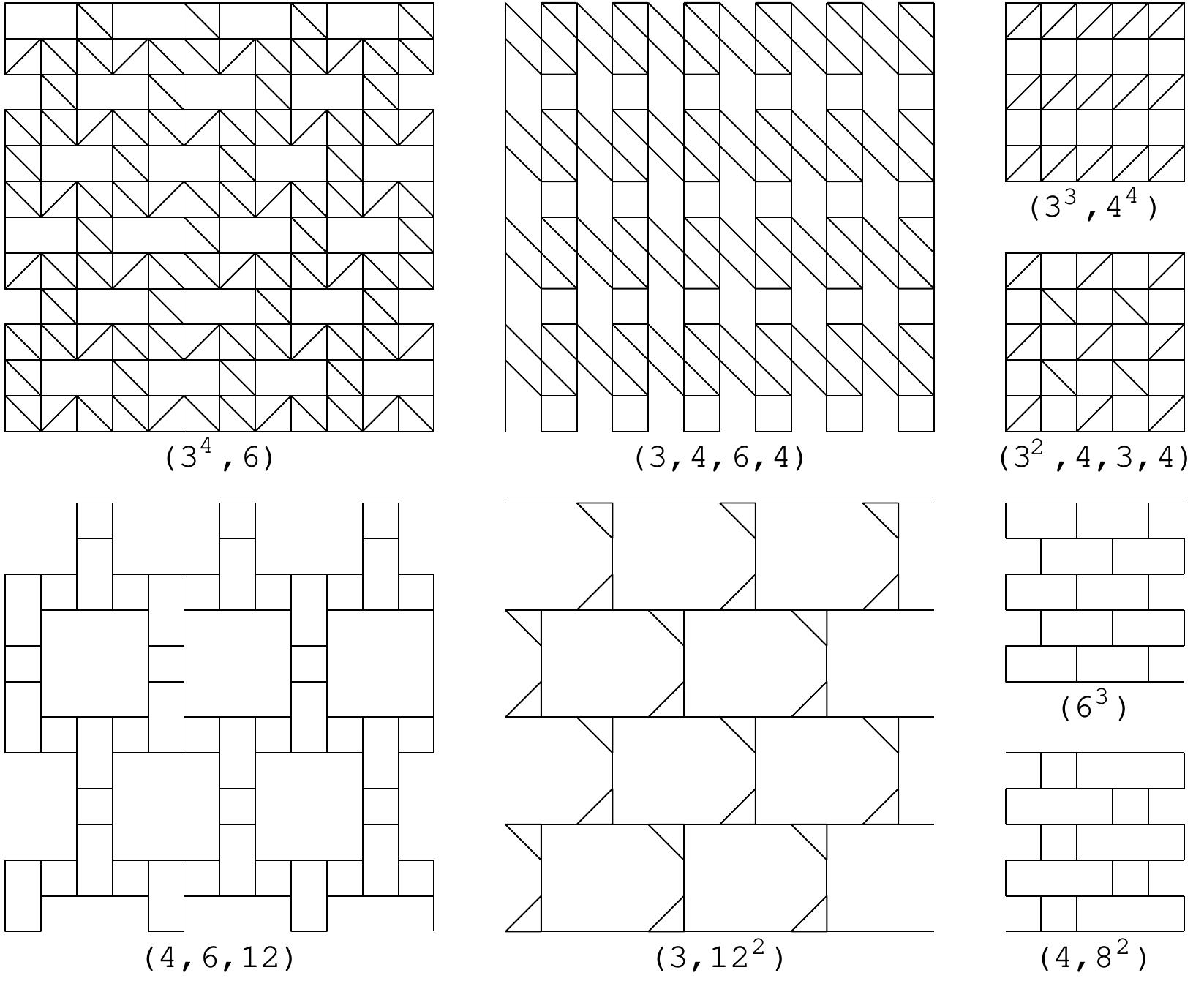}
    \caption{Finite subgraphs of square embeddings of 8 Archimedean lattices.} 
    \label{f:sq}
  \end{center}
\end{figure}

Square representations of the Archimedean lattices studied here, are shown in Figure \ref{f:sq}.

The bond percolation threshold is known exactly for the square, triangular,
and hexagonal lattices (the values are 0.5, $2\sin(\frac{\pi}{18})$,
and $1-2\sin(\frac{\pi}{18})$), and the threshold for the Kagom\'e
lattice has previously been estimated to 0.524\,4053 by Ziff and Suding,
\cite{ZS97}.  Ziff and Scullard's, \cite{ZS06} conjectured value for $(3,12^{2})$ is $0.740\,421\,178\ldots$, and it is consistent with the estimate given here:  $0.740\,4219\ (8)$ --- the difference is less than one standard deviation (further, there is a small positive bias in the estimate).

In this work we estimate the thresholds for all the Archimedean 
lattices with unknown (and in one case conjectured) threshold except the previously studied Kagom\'e lattice. The hexagonal lattice is used as a benchmark.

\subsection{Percolation}
The bond percolation process is defined as follows. For each edge of a
graph $G$, declare the edge to be open with probability $p$,
independently of all other edges, and closed otherwise. 

For the kind
of graphs studied here, it is well known that there exists a critical
value $p_c(G)$, called the critical probability, or threshold, such that for $p>p_c(G)$, there exists an unique infinite
connected component of open edges, while for $p<p_c(G)$, only finitely
large connected components of open edges exist.

\subsection{Motivation}
One objective of this simulation study is to get an empirical
answer to the following question.
\begin{Q}
  If $G$ and $H$ are two Archimedean lattices, with self-avoiding walk connective
  constants $\mu(G)$ and $\mu(H)$, site percolation
  thresholds $p_c^s(G)$ and $p_c^s(H)$, and bond percolation
  thresholds $p_c^b(G)$ and $p_c^b(H)$, is it true that
  \[
    \mu(G) \leq \mu(H)  \Leftrightarrow
    p_c^s(G) \geq p_c^s(H) \Leftrightarrow 
    p_c^b(G) \geq p_c^b(H).
  \]
\end{Q}

The answer is negative for general two dimensional,
quasi-transitive, planar graphs, as shown by Wierman, \cite{W02},
and Parviainen, \cite{P03}.

Available estimates and rigorous bounds for site percolation
thresholds and connective constants suggest that these two
models give the same order on the Archimedean lattices.

The estimates given in this work suggest, however, that there exist two
pairs of Archimedean lattices for which the bond and site percolation
thresholds are in opposite order, namely the pair $(3,6,3,6)$
(Kagom\'e) and $(3,4,6,4)$ (Ruby), and the pair $(3^3,4^2)$
and $(3^2,4,3,4)$.

\section{The method of Newman and Ziff}
Newman and Ziff, \cite{NZ01}, have developed a fast algorithm for
estimating percolation thresholds (both bond and site); the running
time is nearly \emph{linear} in the system size (the number of
vertices or edges of the subgraph), while still producing accurate
estimates. 

It turns out that it is profitable to modify the subgraphs used, and
use torus shaped regions. We will consider $Q(p)$, the probability
that a cluster wraps around the torus in one direction, but not both.

 If we generate a percolation
process on a subgraph with $M$ edges, by adding edges in random
order, we get $M+1$ different, but dependent, realizations of a
percolation process. These give a rough 
estimate of the function $q(n)$, the probability that a cluster spans the torus
in one direction but not the other, given that there are exactly $n$
open edges.

Assume for the moment that  $q(n)$ is exactly known. The law of
total probability then gives us $Q(p)$:
\begin{equation*}\label{convolution}
Q(p)=\sum_{n=0}^{M}q(n){M \choose n}p^{n}(1-p)^{M-n}.
\end{equation*}
Thus from a single realisation of the Newman--Ziff algorithm we get an
estimate of $Q(p)$ for \emph{all} values of $p$, from 0 to 1.

In our case $p_c$ can be estimated by the value of $p$ at which
$Q(p)$ is maximised, since the probability $Q(p)$ tends to zero both
for $p$ both above and below $p_c$, as the system size grows.

For the method to achieve its impressive running time, it is necessary
to keep track of clusters efficiently. This can be achieved by a
tree based union/find algorithm. For each added edge, we
\emph{find} the clusters to which the endpoints belong. If the
clusters are different, the \emph{union} of the clusters is
calculated. Both steps are rapidly done by representing the set of
clusters as a directed forest. By a small modification of the
union/find algorithm, detection of cluster wrapping is also easy.

We can speed up the execution by delaying the convolution (equation (\ref{convolution})) and maximisation step, by averaging over a batch of, say $m$,  estimates of $q(n)$ and use the average to estimate $Q(p)$ and $p_{c}$. One minor drawback by doing so is that we get fewer samples for estimation the statistical error. Also, the standard error\footnote{the standard error is an estimate of the standard deviation of the error of the estimate --- if the estimator is a mean of $n$ values, the standard error is $s.e.=s/\sqrt{n}$ where $s$ is the sample standard deviation.}  does not decrease as $1/\sqrt{m}$ --- but this is only true for small $m$, and a $1/\sqrt{m}$ factor  dominates from $m\approx 10$ an onwards. For the largest system sizes considered here, the standard error for large $m$ is approximately $2/\sqrt{m}$ times that for $m=1$. We thus lose only a factor 2 which is cheap considering that the convolution and maximisation step can be of an order of 100 times more time consuming than the generation step.

We have also run short simulations with the hull gradient method (see for example
\cite{SZ99}), using different representations (of the covering graphs) of
the lattices, to verify the implementations. (A reason
for not using the hull gradient for longer runs was the memory
requirement; the simulations were run on standard desktop computers,
simultaneously used in daily work by the author's colleagues.)

\section{Numerical results}
The finite size error decreases very fast. For our benchmark case, the
hexagonal lattice, the finite size bias was of order
$10^{-8}$ for systems of size 50\,000.

This prompted us to concentrate the computations on only one large
system per lattice. We used square shaped subgraphs, with between
60\,000 and 75\,000 edges, for which we believe the finite size errors to be
of order $10^{-8}$, an order of magnitude smaller than the statistical
errors. The results are summarized in Table \ref{t:res}. The standard error given is $s_{m}/\sqrt{m}$ where $s_{m}$ is the sample standard deviation when $m$ iterations are used for each maximisation step. If we do $n$ realisations, $s_{m}$ is given by
\[s^{2}_{m}=\frac{1}{N-1}\sum_{i=1}^{n}(\hat{p}_{c}^{(i)}-\hat{p}_{c})^{2},\]
where $\hat p_{c}^{(i)}$ is the estimate from realisation $i$ and $\hat{p}_{c}$ is the grand estimate, the mean of the $\hat{p}^{(i)}_{c}$. Approximate $(1-\alpha)\%$ confidence intervals are given by 
\[\hat{p}_{c}\pm z_{\alpha/2}\ s.e.,\]
where $z_{\alpha/2}$ is the $\alpha/2$ quantile of the Gaussian distribution.\footnote{Strictly speaking $t$-distribution quantiles should be used, but for the values of $n$ used here they agree with the Gaussian quantiles.}
 For all
lattices we used batches of $10^5$ realisations for each maximisation step.

In \cite{NZ01}, it was observed that the  standard deviation scales as
\begin{equation*}
  \label{eq:nz-s-scale}
  \sigma\sim M^{-3/8},
\end{equation*}
where $M$ is the system size (the number of edges). In our simulations we generally observe slightly slower convergence. Fitting $\sigma=C M^{-a}$ we found some variations in the estimated values of  $a$ between lattices; we observed values in the range 0.25 to 0.40. The values decrease as the number of iterations $m$ used per maximisation step increases, but appear to settle down for $m\approx 20$. The between-lattice variation also decreases with $m$. As the purpose here is to get precise estimates for a number of lattices, we have not studied the convergence rate thoroughly for all combinations of lattices and parameters. In Table 1 we report estimates of $a$ for $m=100$. For the hexagonal lattice, Figure 2 shows a log-log-log plot of the sample standard deviation as a function of system size and number of iterations per maximisation step. The numbers in parenthesis give the estimated values of the exponents $a$ for fixed values of $m$. 

\begin{figure}[htbp]
  \begin{center}
    \includegraphics[scale=1.4]{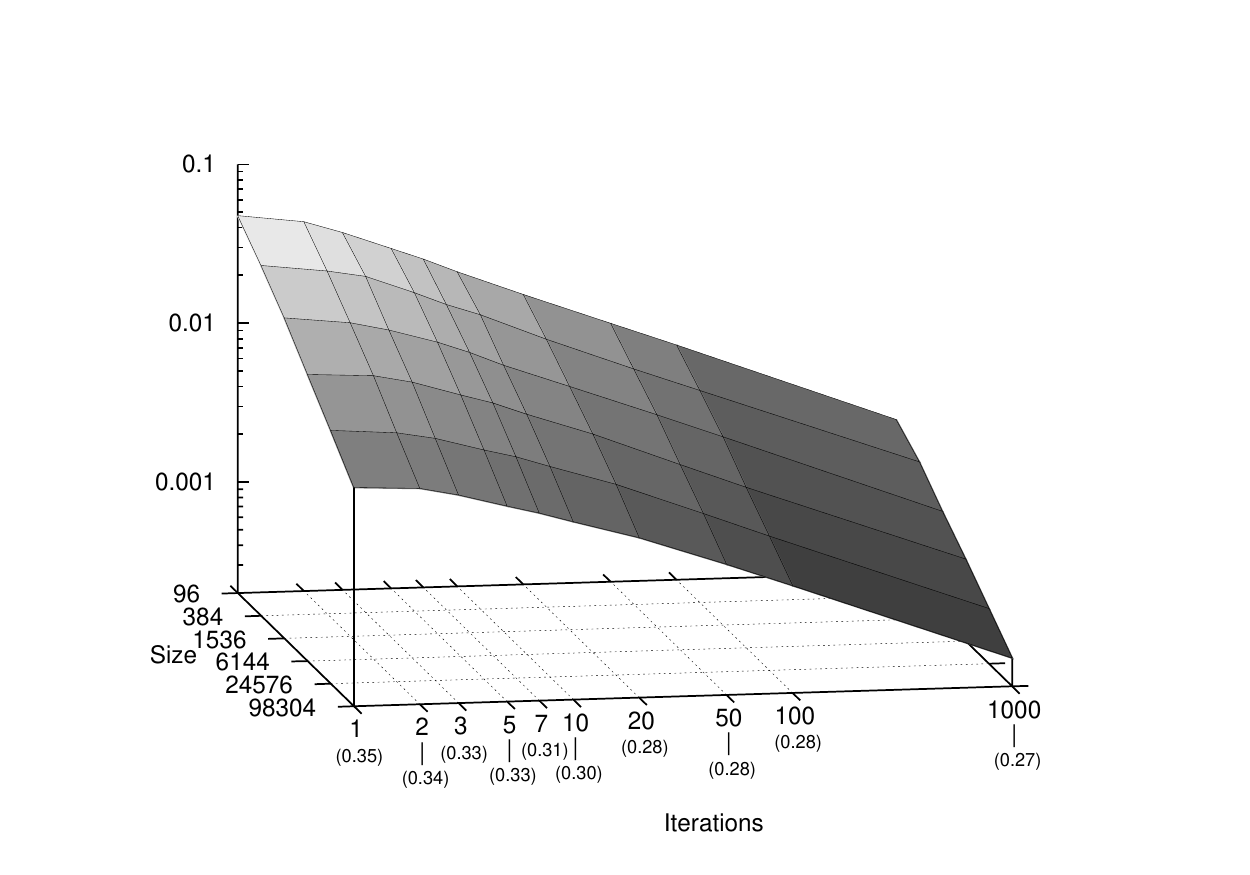}
    \caption{Sample standard deviation as function of system size and number of iterations per maximisation step.} 
    \label{f:ssd}
  \end{center}
\end{figure}

\begin{table}[htbp]
  \caption{Simulation results. The standard error is denoted by s.e., the number of realizations by $n$, the system size by $M$, 
  and the estimated value of the exponent $a$ is denoted by $\hat a$.    } 
  \label{t:res}
  \begin{center} 
 \begin{tabular}{lcccccc}
      Lattice & $\hat{p}_c$ & s.e.$/10^{-7}$ & $n$ & $M$&$\hat{a}$\\
	  \hline	
      $(3,12^2)$    & 0.740\,421\,95 &8.0&2485&67\,500&0.276\\
      $(4,6,12)$    & 0.693\,733\,83 &7.2&2930&73\,728&0.277\\
      $(4,8^2)$     & 0.676\,802\,32 &6.3&2983&60\,000&0.284\\
      $(3,4,6,4)$   & 0.524\,832\,58 &5.3&7568&64\,800&0.286\\
      $(3^4,6)$     & 0.434\,306\,21 &5.0&7397&64\,000&0.274\\
      $(3^3,4^2)$   & 0.419\,641\,91 &4.3&9822&64\,000&0.274\\
      $(3^2,4,3,4)$ & 0.414\,137\,43 &4.6&7504&64\,000&0.269\\
    \end{tabular}    
  \end{center}
\end{table}

 For the hexagonal lattice we did long simulations for 6 different
system sizes to study the precision. We used batches of $10^{4}$ realisations. The results are summarized in Table \ref{t:hex}. The exact value
of the bond percolation threshold is $p_c=1-2\sin(\pi/18)\approx 0.652\,703\,6446$.

\begin{table}[htbp]
  \caption{Simulation results for the hexagonal lattice. Here $M$ denotes the number of edges, $n$ the total number of realizations used for the estimate $\hat p_c$, with standard error s.e. Further, $s_{1}$ is the sample standard deviation when one repetition is used for each maximisation step, and $s_{\infty}$ is an estimate of the limiting sample standard deviation per repetition when $m$ repetitions are used for each maximisation step.} 
  \label{t:hex}
  \begin{center}
  \begin{tabular}{rrccccc}
      $M$&$n/10^4$&$\hat p_c$&s.e.&$s_{1}$&$s_{\infty}$\\
	\hline
      96&754\,700&0.652\,685\,19 &$6.68\times 10^{-7}$ &0.048&0.058\\
      384&634\,300 &0.652\,707\,81 &$5.38\times 10^{-7}$ &0.032&0.043\\
      1536&460\,520  &0.652\,704\,77 &$4.49\times 10^{-7}$ &0.021&0.030\\
      6144&444\,786  &0.652\,703\,77 &$3.12\times 10^{-7}$ &0.013&0.021\\
      24\,576&41\,727  &0.652\,703\,87 &$6.58\times 10^{-7}$ &0.0079&0.014\\
      98\,304&\,5326&0.652\,703\,67  &$1.32\times 10^{-6}$ &0.0048&0.0097\\
    \end{tabular}
    \end{center}
\end{table}

 The small finite size bias observed in \cite{NZ01} is confirmed by
our simulations. The estimates $\hat p_{c}$ are conjectured to converge like 
\[|\hat p_{c}-p_{c}|\sim M^{-11/8}.\]
Fitting 
\[\hat p_{c}=p_{c}+b M^{-11/8}\]
to data, excluding the $M=96$ data point, gives $b=0.0017$ and 
a finite size error for $M=50\,000$ of $3.4\times 10^{-8}$.

Excluding further data points for small values of $M$ gives slight variations, but the finite size error is always below $4\times 10^{8}$. This supports our belief that the finite size error for our main estimates are
considerable less than the statistical error.

It is also worthwhile to note that, except for very small sizes,  the finite size bias is
positive, in contrast to the site percolation case studied in
\cite{NZ01}. (The same is observed with the hull gradient method; for the bond
cases studied here, the finite size bias is positive, while the site
cases on the same lattices studied in \cite{SZ99} have negative
finite size biases.)

\section*{Acknowledgement}
I am grateful to Robert Ziff, for many helpful answers and comments and for encouragement. I also acknowledge the support received from the Australian Research Council Centre of Excellence for Mathematics and Statistics of Complex Systems (MASCOS).

\end{document}